\documentclass{article}

\PassOptionsToPackage{numbers, compress}{natbib}

     \usepackage[final]{neurips_2019}    

\usepackage[utf8]{inputenc} 
\usepackage[T1]{fontenc}    
\usepackage{hyperref}       
\usepackage{url}            
\usepackage{booktabs}       
\usepackage{amsfonts}       
\usepackage{nicefrac}       
\usepackage{microtype}      
\usepackage{soul} 
\usepackage{graphicx} 
\usepackage{upgreek}

\title{Biophysical models of cis-regulation as\\ interpretable neural networks}

\author{
  Ammar Tareen \\
  Simons Center for Quantitative Biology\\
  Cold Spring Harbor Laboratory\\
  Cold Spring Harbor, NY 11724 \\
  \texttt{tareen@cshl.edu} \\
  \And
    Justin B. Kinney \\
  Simons Center for Quantitative Biology\\
  Cold Spring Harbor Laboratory\\
  Cold Spring Harbor, NY 11724 \\
  \texttt{jkinney@cshl.edu} 
}

\begin{document}

\maketitle

\begin{abstract}

The adoption of deep learning techniques in genomics has been hindered by the difficulty of mechanistically interpreting the models that these techniques produce. In recent years, a variety of post-hoc attribution methods have been proposed for addressing this neural network interpretability problem in the context of gene regulation. Here we describe a complementary way of approaching this problem. Our strategy is based on the observation that two large classes of biophysical models of cis-regulatory mechanisms can be expressed as deep neural networks in which nodes and weights have explicit physiochemical interpretations. We also demonstrate how such biophysical networks can be rapidly inferred, using modern deep learning frameworks, from the data produced by certain types of massively parallel reporter assays (MPRAs). These results suggest a scalable strategy for using MPRAs to systematically characterize the biophysical basis of gene regulation in a wide range of biological contexts. They also highlight gene regulation as a promising venue for the development of scientifically interpretable approaches to deep learning. 

\end{abstract}


Deep learning -- the use of large multi-layer neural networks in machine learning applications -- is revolutionizing information technology \cite{lecun_2015}. There is currently a great deal of interest in applying deep learning techniques to problems in genomics, especially for understanding gene regulation \cite{zhou_troyanskaya, alipanahi_deepbind, Jaganathan:2019fv, eraslan}. These applications remain somewhat controversial, however, due to the difficulty of mechanistically interpreting neural network models trained on functional genomics data. Multiple attribution strategies, which seek to extract meaning post-hoc from neural networks that have  rather generic architectures, have been proposed for addressing this  interpretability problem \cite{simonyan_2013, deeplift, causal_attribution}. However, there remains a substantial gap between the outputs of such attribution methods and fully mechanistic models of gene regulation. 

Here we advocate for a complementary approach: the inference of neural network models whose architecture reflects explicit biophysical hypotheses for how cis-regulatory sequences function. This strategy is based on two key observations. First, two large classes of biophysical models can be formulated as three-layer neural networks that have a stereotyped form and in which nodes and weights have explicit physiochemical interpretations. This is true of thermodynamic models, which rely on a quasi-equilibrium assumption \cite{Ackers:1982tq, Shea:1985tz, Bintu:2005bn, Bintu:2005ur, Segal:2009jv, Sherman:2012gu}, as well as kinetic models, which are more complex but do not make such assumptions \cite{Estrada:2016ct, Scholes:2017gz, Park:2019dx}. Second, existing deep learning frameworks such as TensorFlow \cite{abadi2016tensorflow} are able to rapidly infer such models from the data produced by certain classes of MPRAs.  

The idea of using neural networks to model the biophysics of gene regulation goes back to at least \cite{mjolness}. To our knowledge, however, the fact that all thermodynamic models of gene regulation can be represented by a simple three-layer architecture has not been previously reported. We are not aware of any prior work that uses neural networks to represent kinetic models or models involving King-Altman diagrams. There is a growing literature on modeling MPRA data using neural networks \cite{Rosenberg:2015em, Cuperus:2017cq, Movva:2019ea, Sample:2019km, Bogard:2019cl}. However, such modeling has yet to be advocated for MPRAs that dissect the mechanisms of single cis-regulatory sequences, such as in \cite{Patwardhan:2009cw, Kinney:2010tn, Melnikov:2012dw, Kwasnieski:2012hu, Patwardhan:2012hy}, which is our focus here. Finally, some recent work has used deep learning frameworks to infer parametric models of cis-regulation \cite{Liu:2019br, deBoer:2017cj}. These studies did not, however, infer the type of explicit biophysical quantities, such as $\Delta G$ values, that our approach recovers. 


\begin{figure}[t]
    \centering
    \includegraphics{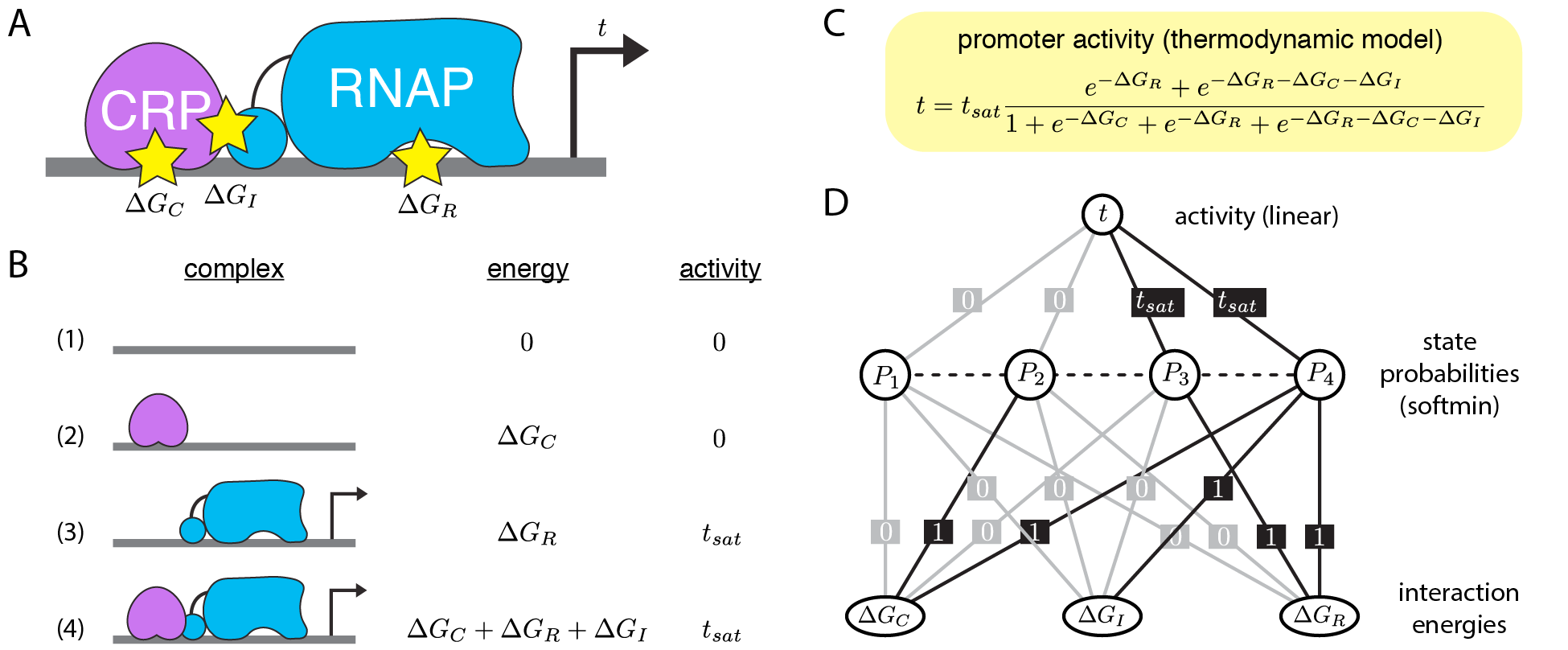}
    \caption{A thermodynamic model of transcriptional regulation. (A) Transcriptional activation at the \textit{E.\ coli} \textit{lac} promoter is regulated by two proteins, CRP and $\upsigma^{70}$ RNA polymerase (RNAP). CRP is a transcriptional activator that up-regulates transcription by stabilizing RNAP-DNA binding. $\Delta G_C$ and $\Delta G_R$ respectively denote the Gibbs free energies of the CRP-DNA and RNAP-DNA interactions, while $\Delta G_I$ denotes the Gibbs free energy of interaction between CRP and RNAP. (B) Like all thermodynamic models of gene regulation, this model consists of a set of states, each state having an associated Gibbs free energy and  activity. The probability of each state is assumed to follow the Boltzmann distribution. (C) The corresponding activity predicted by such thermodynamic models is the state-specific activity averaged together using these Boltzmann probabilities.  (D) This model formulated as a three-layer neural network. First layer nodes represent interaction energies, second layer nodes represent state probabilities, and third layer nodes represent transcriptional activity. The values of weights are indicated; gray lines correspond to zero weights. The second layer has a softmin activation, while the third has a linear activation. All thermodynamic models of cis-regulation can be represented using this general three-layer form.}
    \label{fig:thermo}
\end{figure}

All thermodynamic models of cis-regulation can be represented as three-layer neural networks as follows. First one defines a set of molecular complexes, or ``states'',  which we index using $s$. Each state has both a Gibbs free energy $\Delta G_s$ and an associated activity $\alpha_s$. These energies determine the probability $P_s$ of each state occurring in thermodynamic equilibrium via the Boltzmann distribution,\footnote{To reduce notational burden, all $\Delta G$ values are assumed to be in thermal units. At  $37^\circ$C, one thermal unit is  $1\, k_B T = 0.62~\mathrm{kcal/mol}$, where $k_B$ is Boltzmann's constant and $T$ is temperature.}  
\begin{equation}
    P_s = \frac{e^{-\Delta G_s}}{\sum_{s'} e^{-\Delta G_{s'}}}.
\end{equation}
The energy of each state is, in turn, computed using integral combinations of the individual interaction energies $\Delta G_j$ that occur in that state. We can therefore write $\Delta G_s = \sum_j \omega_{sj} \Delta G_j$, where $\omega_{sj}$ is the number of times that interaction $j$ occurs in state $s$. The resulting activity predicted by the model is given by the activities $\alpha_s$ of the individual states averaged over this distribution, i.e., $t = \sum_s \alpha_s P_s$.

Fig.\ \ref{fig:thermo} illustrates a thermodynamic model for transcriptional activation at the \textit{E.\ coli lac} promoter. This model involves two proteins, CRP and RNAP, as well as three interaction energies: $\Delta G_C$, $\Delta G_R$, and $\Delta G_I$.  The rate of transcription $t$ is further assumed to be proportional to the fraction of time that RNAP is bound to DNA (Fig.\ \ref{fig:thermo}A). This model is summarized by four different states, two of which lead to transcription and two of which do not (Fig.\ \ref{fig:thermo}B). Fig.\ \ref{fig:thermo}C shows the resulting formula for $t$ in terms of model parameters. This model is readily formulated as a feed-forward neural network (Fig.\ \ref{fig:thermo}D).  Indeed, all thermodynamic models of cis-regulation can be formulated as three-layer neural networks: layer 1 represents molecular interaction energies, layer 2 (which uses a softmin activation) represents state probabilities, and layer 3 (using linear activation) represents the biological activity of interest, which in this case is transcription rate. 

\begin{figure}[t]
    \centering
    \includegraphics{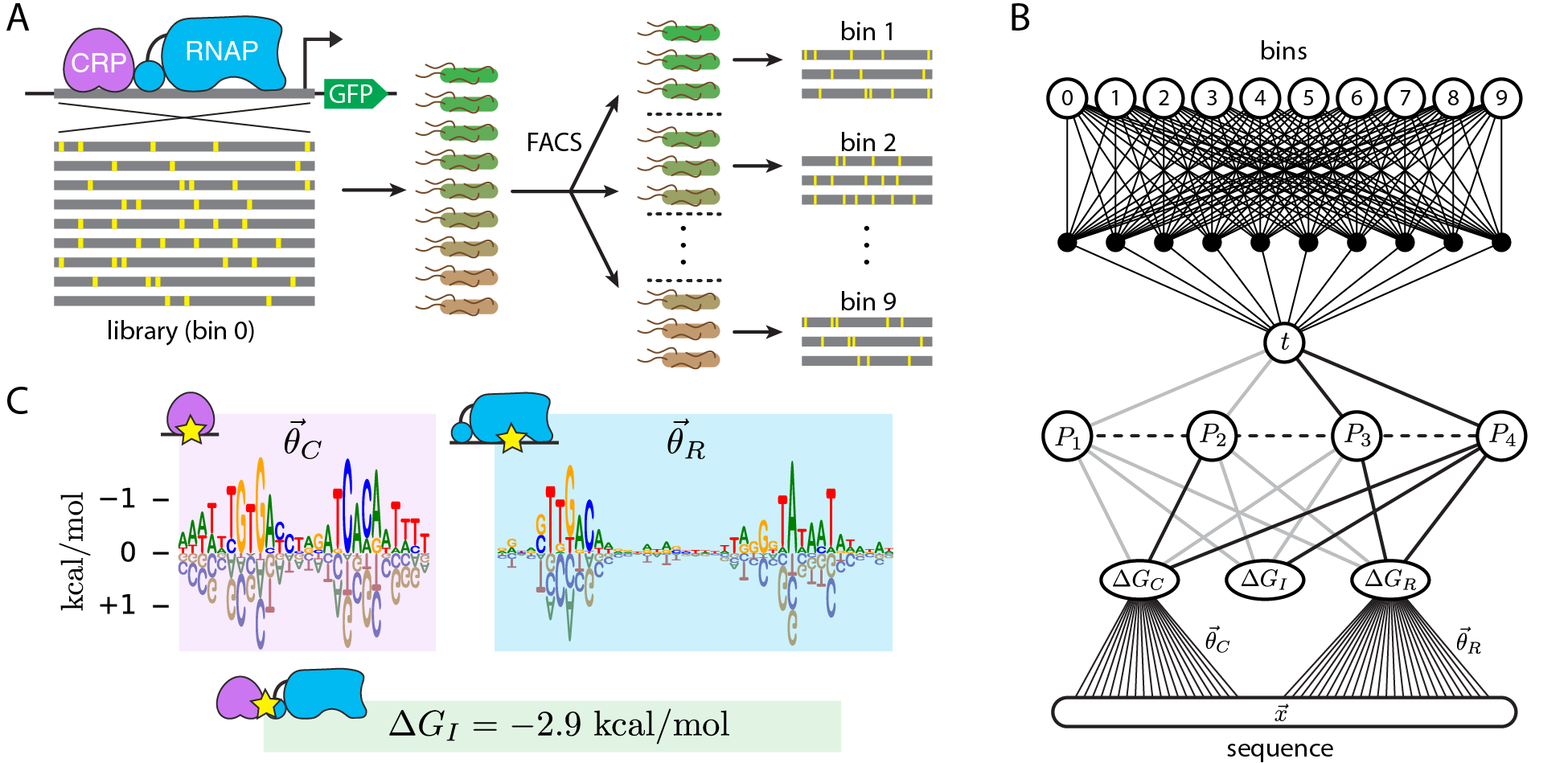}
    \caption{Inference of a thermodynamic model from MPRA data. (A) Schematic of the sort-seq MPRA of \cite{Kinney:2010tn}. A 75 bp region of the \textit{E.\ coli lac} promoter was mutagenized at 12\% per nucleotide. Variant promoters were then used to drive the expression of GFP. Cells carrying these expression constructs were then sorted using FACS, and the variant sequences in each bin were sequenced. This yielded data on about $5 \times 10^4$ variant promoters across 10 bins. (B) The neural network from Fig.\ \ref{fig:thermo}D, but with $\Delta G_C$ and $\Delta G_R$ expressed as linear functions of the DNA sequence $\vec{x}$, as well as a dense feed-forward network mapping activity $t$ to bins via a probability distribution $p(\mathrm{bin}|t)$. Gray lines indicate weights fixed at 0. The weights linking nodes $P_3$ and $P_4$ to node $t$ were constrained to have the same value $t_{sat}$. (C) The parameter values inferred from the MPRA data of \cite{Kinney:2010tn}. Shown are the CRP energy matrix $\vec{\theta}_C$, the RNAP energy matrix $\vec{\theta}_R$, and the CRP-RNAP interaction energy $\Delta G_I$. Since increasingly negative energy corresponds to stronger binding, they $y$-axis in the logo plots is inverted. Logos were generated using Logomaker \cite{Tareen:2019cq}.}
    \label{fig:mpra}
\end{figure}


We can infer thermodynamic models like these for a cis-regulatory sequence of interest (the wild-type sequence) from the data produced by an MPRA performed on an appropriate sequence library \cite{Kinney:2010tn}. Indeed, a number of MPRAs have been performed with this explicit purpose in mind \cite{Kinney:2010tn, Melnikov:2012dw, RazoMejia:2014fo, Belliveau:2018kr, Barnes:2019hxa}. To this end, such MPRAs are generally performed using libraries that consist of sequence variants that differ from the wild-type sequence by a small number of single nucleotide polymorphisms (SNPs). The key modeling assumption that motivates using libraries of this form is that the assayed sequence variants will form the same molecular complexes as the wild-type sequence, but with Gibbs free energies and state activities whose values vary from sequence to sequence. By contrast, variant libraries that contain insertions, deletions, or large regions of random DNA (e.g.\ \cite{Rosenberg:2015em, Cuperus:2017cq, Sample:2019km, Bogard:2019cl, deBoer:2017cj}) are unlikely to satisfy this modeling assumption.

Fig.\ \ref{fig:mpra}A summarizes the sort-seq MPRA described in \cite{Kinney:2010tn}. \textit{Lac} promoter variants were used to drive GFP expression in \textit{E.\ coli}, cells were sorted into 10 bins using fluorescence-activated cell sorting, and the variant promoters within each bin were sequenced. This yielded data comprising about $5 \times 10^4$ variant \textit{lac} promoter sequences, each associated with one of 10 bins. The authors then fit the biophysical model shown in Fig.\ \ref{fig:thermo}C, but under the assumption that $\Delta G_C = \vec{\theta}_C \cdot \vec{x} + \mu_C$ and $\Delta G_R = \vec{\theta}_R \cdot \vec{x} + \mu_R$, where $\vec{x}$ is a one-hot encoding of promoter DNA sequence. 

Here we used TensorFlow to infer the same model formulated as a deep neural network. Specifically, we augmented the network in Fig. \ref{fig:thermo}D by making $\Delta G_C$ and $\Delta G_R$ sequence-dependent as in \cite{Kinney:2010tn}. To link $t$ to the MPRA measurements, we introduced a feed-forward network with one hidden layer and a softmax output layer corresponding to the 10 bins into which cells were sorted. Model parameters were then fit to the MPRA dataset using stochastic gradient descent and early stopping. The results agreed well with those reported in \cite{Kinney:2010tn}. In particular, the parameters in the energy matrices for CRP ($\vec{\theta}_C$) and RNAP ($\vec{\theta}_R$) exhibited respective Pearson correlation coefficients of 0.986 and 0.994 with those reported in  \cite{Kinney:2010tn}. The protein-protein interaction energy that we found, $ \Delta G_I = -2.9$ kcal/mol, was also compatible with the previously reported value  $\Delta G_I =-3.3 \pm 0.4 $ kcal/mol. 

A major difference between our results and those of \cite{Kinney:2010tn} is the ease with which they were obtained. Training of the network in Fig.\ \ref{fig:mpra}B consistently took about 15 minutes on a standard laptop computer. The model fitting procedure in \cite{Kinney:2010tn}, by contrast, relied on a custom Parallel Tempering Monte Carlo algorithm that took about a week to run on a multi-node computer cluster (personal communication), and more recent efforts to train biophysical models on MPRA data have encountered similar computational bottlenecks \cite{Belliveau:2018kr, Barnes:2019hxa}.  

Also of note is the fact that in \cite{Kinney:2010tn} the authors inferred models using information maximization. Specifically, the authors fit the parameters of $t(\vec{x})$ by maximizing the mutual information $I[t; \mathrm{bin}]$ between model predictions and observed bins. One  difficulty with this strategy is the need to estimate mutual information. Instead, we used maximum likelihood to infer the parameters of $t(\vec{x})$ as well as the experimental transfer function (i.e., noise model) $p(\mathrm{bin}|t)$, which was modeled by a dense feed-forward network with one hidden layer. These two inference methods, however, are essentially equivalent: in the large data regime, the parameters of $t$ that maximize $I[t; \mathrm{bin}]$ are the same as the parameters one obtains when maximizing likelihood over the parameters of \textit{both} $t$ and $p(\mathrm{bin} | t)$; see \cite{Kinney:2007dh, Kinney:2014ge, Atwal:2015dj}. 

%
%

\begin{figure}[t]
    \centering
    \includegraphics{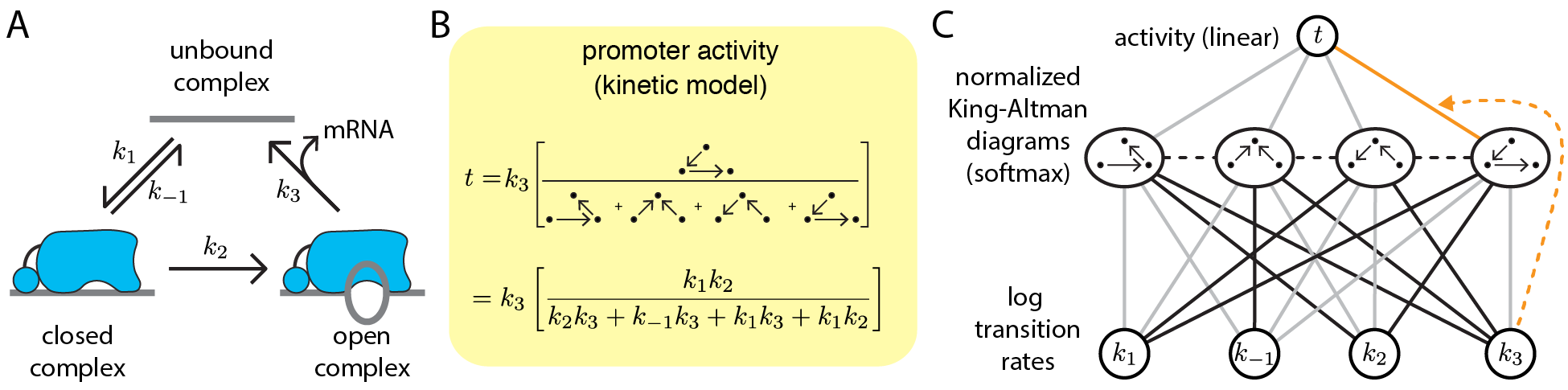}
    \caption{A kinetic model for transcriptional initiation by \textit{E.\ coli} RNAP. (A) In this model, promoter DNA can participate in three complexes: unbound, closed, and open \cite{McClure:1980wy, McClure:1985cy}. Transitions between these complexes are governed by four rate constants: $k_1$, $k_{-1}$, $k_2$, and $k_3$. (B) A formula for the steady-state rate of mRNA production can be obtained using King-Altman diagrams \cite{King:1956vf, Hill:1989te}. (C) This formula can be represented using the three-layer neural network shown, where layer 1 represents log transition rates, layer 2 (softmax activation) represents normalized King-Altman diagrams, and layer 3 (linear activation) represents promoter activity. Black lines indicate weight 1; gray lines indicate weight 0. Note that the single nonzero weight connecting layer 2 to layer 3 (orange) is actually the transition rate $k_3$ from layer 1. All kinetic models of cis-regulation share this general three-layer form.}
    \label{fig:kinetic}
\end{figure}

A shortcoming of thermodynamic models is that they ignore non-equilibrium processes. Kinetic models address this problem by providing a fully non-equilibrium characterization of steady-state activity. Kinetic models are specified by listing explicit state-to-state transition rates rather than Gibbs free energies. Fig.\ \ref{fig:kinetic}A shows a three-state kinetic model of transcriptional initiation consisting of unbound promoter DNA, an RNAP-DNA complex in the closed conformation, and the RNAP-DNA complex in the open conformation \cite{McClure:1980wy, McClure:1985cy}. The rate $k_3$ going from the open state to the unbound state represents transcript initiation. The overall transcription rate in steady state is therefore $k_3$ times the occupancy of the open complex. 

King-Altman diagrams \cite{King:1956vf, Hill:1989te}, a technique from mathematical enzymology, provide a straight-forward way to compute steady-state occupancy in kinetic models. Specifically, each state's occupancy is proportional to the sum of directed spanning trees (a.k.a.\ King-Altman diagrams) that flow to that state, where each spanning tree's value is given by the product of rates comprising that tree. Fig.\ \ref{fig:kinetic}B illustrates this procedure for the kinetic model in Fig.\ \ref{fig:kinetic}A. Every such kinetic model can be  represented by a three-layer neural network (e.g., Fig.\ \ref{fig:kinetic}C) in which first layer nodes represent log transition rates, second layer nodes (after a softmax activation) represent normalized King-Altman diagrams, and third layer nodes represent the activities of interest.

Here we have shown how both thermodynamic and kinetic models of gene regulation can be formulated as three-layer deep neural networks in which nodes and weights have explicit biophysical meaning. This represents a new strategy for interpretable deep learning in the study of gene regulation, one complementary to existing post-hoc attribution methods. We have further demonstrated that a neural-network-based thermodynamic model can be rapidly inferred from MPRA data using TensorFlow. This was done in the context of a well-characterized bacterial promoter because previous studies of this system have established concrete results against which we could compare our inferred model. The same modeling approach, however, should be readily applicable to a wide variety of biological systems amenable to MPRAs, including transcriptional regulation and alternative mRNA splicing in higher eukaryotes. 


\noindent\textbf{Code Availability and Acknowledgements}: The neural network model shown in Fig.\ 2C, as well as the scripts used to infer it from the data of \cite{Kinney:2010tn}, are available at \url{https://github.com/jbkinney/19_mlcb}. We thank Anand Murugan, Yifei Huang, Peter Koo, Alan Moses, and Mahdi Kooshkbaghi for helpful discussions, as well as and three anonymous referees for providing constructive criticism. This project was supported by NIH grant 1R35GM133777 and a grant from the CSHL/Northwell Health partnership.

\bibliographystyle{ieeetr}

\bibliography{19_mclb}

\end{document}